\documentclass[final,1p,times]{elsarticle} 
\usepackage{graphicx} 
\usepackage{amssymb} 
\usepackage{amsthm} 

\journal{Nuclear Physics A} 
\begin{document} 

\begin{frontmatter} 


\title{Forward-rapidity azimuthal and radial 
flow of identified particles \\ for 
$\sqrt{s_{NN}}$ = 200 GeV Au+Au collisions}

\author{S.J. Sanders for the BRAHMS Collaboration}

\address{University of Kansas, Lawrence, KS 66045}

\begin{abstract} 
A strong azimuthal flow signature at RHIC suggests rapid system equilibration 
leading to an almost perfect fluid state. The longitudinal extent of the flow 
behavior depends on how this state is formed and can be studied 
by measuring the pseudorapidity and transverse momentum 
dependence of the second Fourier component 
($v_{2}(p_{T})$) of the azimuthal angular distribution. We report on a measurement 
of identified-particle $v_{2}$ as a function of $p_{T}$ (0.5-2.0 GeV/c), 
centrality (0-25$\%$, 25-50$\%$), and pseudorapidity 
($0\leq\eta<3.2$) for $\sqrt{s_{NN}} = 200~\rm GeV$ Au+Au collisions. 
The BRAHMS spectrometers are used for particle identification ($\pi$, K, p) 
and momentum determination and the BRAHMS global detectors 
are used to determine the corresponding  reaction-plane angles. 
The results are discussed in terms of the 
pseudorapidity dependence of constituent quark scaling and in terms of
models that develop  the complete (azimuthal and radial) hydrodynamic 
aspects of the forward dynamics at RHIC.
\end{abstract} 

\end{frontmatter} 



In a non-central collision of two relativistic heavy-ions, 
strong pressure gradients are set up in the almond-shaped 
interaction region.  Early results from RHIC have shown that this overlap 
region behaves as an almost perfect fluid, with the greater pressure that 
exists at the waist of the almond leading to greater particle production 
near the reaction plane.  This azimuthal asymmetry in particle production 
is characterized by the strength of the second Fourier component ($v_{2}$) of 
a harmonic expansion of the angular distribution.  The outward pressure 
can also lead to an outward radial flow of the streaming particles, giving 
a velocity boost to these particles.  To fully understand the hydrodynamic 
properties of RHIC collisions, it is necessary to determine the integral 
and differential $v_{2}$ behavior, as well as establish 
the particle spectra and relative particle
yields~\cite{hirano06}.   
By exploring the interplay of elliptic and radial flow at both mid- and 
forward pseudorapidity, the BRAHMS results better constrain models of the initial 
conditions and longitudinal extent of the interaction region.    

\begin{figure}[!ht]
\begin{center}
\resizebox{!}{6cm}
{\includegraphics{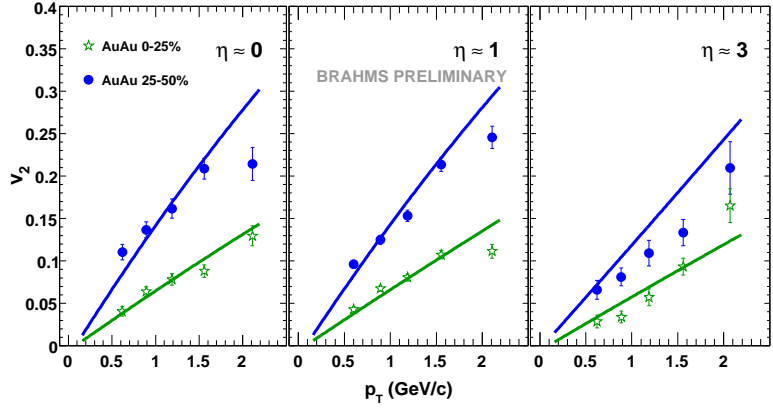}}
\end{center}
\caption{\label{fig:fig1} Charge-hadron $v_{2}(p_T)$ for central and mid-central events 
for $\eta\approx$ 0, 1 and 3. The curves show 3D Hydro+Cascade calculations~\cite{hirano04}. } 
\end{figure}

The measurement explores the correlation of charged hadrons and identified 
particles detected in the two BRAHMS spectrometers with respect to 
reaction planes deduced using four azimuthally symmetric rings of detectors 
arranged around the beam line as part of the experiment's multiplicity 
array~\cite{brahms}.   
The  $v_{2}(p_T)$ dependence of particles 
detected in the spectrometers was determined by the standard reaction-plane 
method~\cite{poskanzer}.  The reaction-plane resolution correction 
was based on a full GEANT simulation of the BRAHMS 
experimental response.  A pseudoevent generator was used to obtain a 
particle throw consistent with previously established particle spectra 
measured using the BRAHMS spectrometers, with the azimuthal asymmetry of 
the particle throw set to reproduce the PHOBOS integral $v_2$ results~\cite{phobos}. 
The correction factor was taken as the ratio of the $v_2(p_T)$ values obtained
from the reconstructed events with that input to the event generator.    

Figure \ref{fig:fig1} shows the resulting charged-hadron 
$v_{2}(p_T)$ values for 0-25$\%$ and 25-50$\%$ central events. The increasing $v_{2}$
behavior up to $p_T\approx 1.5~{\rm GeV/c}$ is characteristic of hydrodynamic 
flow~\cite{heiselberg}.
Near mid-rapidity
the results show very little pseudorapidity dependence. In going to forward pseudorapidity 
with $\eta\approx$ 3, the slope is found to decrease for both centrality selections, with a 
greater decrease for the mid-central events.   
The curves show the results of 3D Hydro+Cascade calculations
employing Glauber motivated initial conditions~\cite{hirano04}. Good agreement is found with
experiment for all but the mid-central, forward-pseudorapidity results.  Turning off the hadronic 
cascade part of the calculation leads to near pseudorapidity independence of the $v_{2}(p_T)$
results, in strong disagreement with the observed slopes 
at $\eta\approx$ 3. Folding the Fig. \ref{fig:fig1} results with the corresponding 
charged hadron spectra measured at $\eta\approx$ 0 and 3 yield
integral v$_{2}$ values of $0.036\pm 0.005$ and $0.027\pm 0.004$, respectively, at the two 
pseudorapidities, in good agreement with the PHOBOS integral $v_2$ results~\cite{phobos}.   

\begin{figure}[!ht]
\begin{center}
\resizebox{!}{6cm}
{\includegraphics{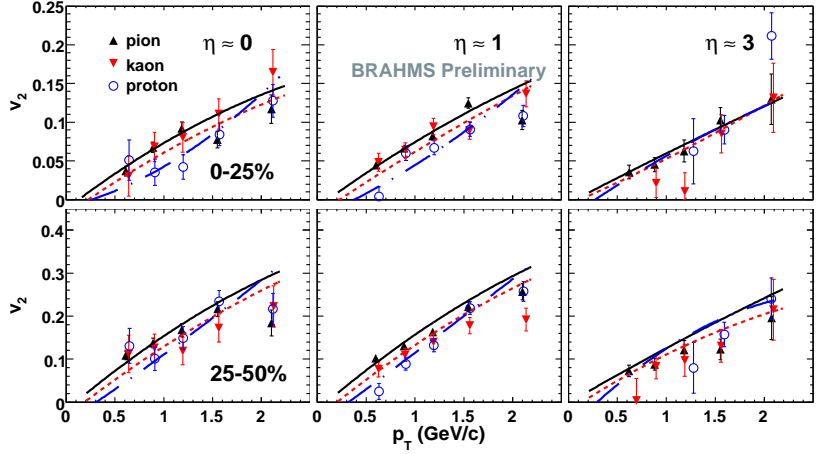}}
\end{center}
\caption{\label{fig:fig2} Identified particle $v_{2}(p_T)$ for central and 
mid-central events for $\eta\approx$ 0, 1 and 3. The curves show 3D Hydro+Cascade 
calculations~\cite{hirano04} for pions(solid),  kaons(short dash), and protons (long dash).} 
\end{figure}

The identified particle $v_2(p_T)$ results are shown in Fig.~\ref{fig:fig2}.  Again, the 
3D Hydro+Cascade calculations, as shown by the smooth curves, 
are in good agreement with experiment
except for the forward-pseudorapidity, mid-central events.  
In addition to reproducing the experimental 
slopes, the calculations also do a good job reproducing the observed mass ordering.     

\begin{figure}[!ht]
\begin{center}
\resizebox{!}{6cm}
{\includegraphics{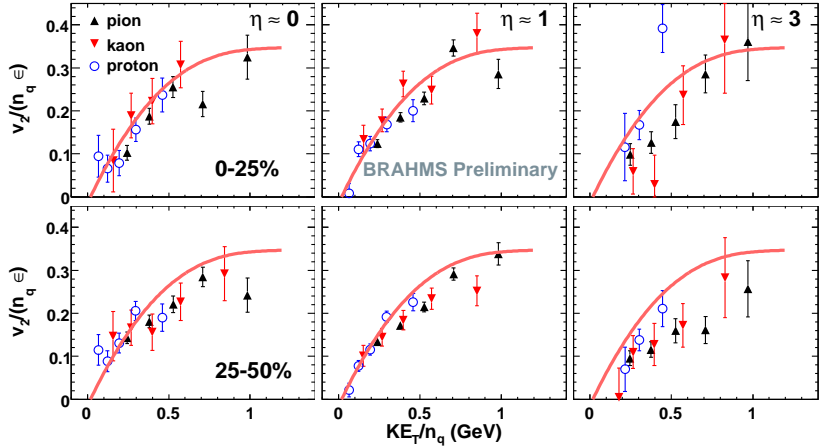}}
\end{center}
\caption{\label{fig:fig3} $v_2$ scaled by the number of valence quarks $n_q$ and the
participant eccentricity $\epsilon$ as a function of the transverse 
kinetic energy KE$_{\rm{T}}$ scaled by $n_q$. The curve, common to all panels, is based on the
systematics observed for a large number particle types at RHIC~\cite{lacey}.} 
\end{figure}

One of the remarkable features of the RHIC elliptic flow behavior is how closely it follows that
expected for a perfect fluid.  This is particularly evident when the elliptic 
flow $v_2(p_T)$
values are scaled by the eccentricity $\epsilon$ of the overlap region and the 
number of consituent quarks
of the detected particle $n_q$ and then plotted against the mean transverse energy 
per constituent quark, $\langle E_T\rangle /n_q$, as
shown in Fig.~\ref{fig:fig3} for the BRAHMS results at $\eta\approx $0, 1, and 3.
With this scaling, the elliptic flow observed for $\sqrt{s_{\rm{NN}}}=200$~GeV Au+Au collisions 
for a large number of different outgoing particle types are found to follow a common 
trend~\cite{lacey}, 
as shown by the curve in the figure.  Such behavior is consistent with the creation of a 
near-perfect 
fluid, with the constituent quark scaling suggesting this fluid involves quark 
degrees-of-freedom, 
as expected in coalescence models.   
Our central results are consistent with the established systematics at all three 
pseudorapidities.  
For mid-central collisions, the data continue to track well with the systematics 
for $\eta\approx$0 and 1, but
show significantly reduced elliptic flow at forward pseudorapidity. As also found in 
comparison of 3D Hydro+Cascade 
results to our data, the mid-central events at forward pseudorapidity suggest a 
process other than ideal hydrodynamics 
is playing a role.

\begin{figure}[!ht]
\begin{center}
\resizebox{6cm}{!}
{\includegraphics{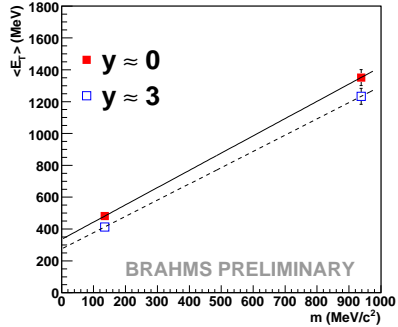}}
\end{center}
\caption{\label{fig:fig4} Transverse kinetic energy $\langle E_T\rangle$ as a function of the mass for pions and 
protons at  rapidity y$\approx$0 and 3. The curves are linear fits to guide the eye.} 
\end{figure}

The elliptic flow behavior is believed to be established at an early stage of the reaction, with the integral 
elliptic flow largely fixed at the point of chemical freeze out~\cite{hirano06}.  
The differential elliptic flow also depends
on the subsequent hadronization stage, where radial flow can significantly affect the final particle spectra. 
Radial flow results in a velocity boost of the outwardly streaming particles.
Figure~\ref{fig:fig4} shows  
the observed $\langle E_T\rangle$ values for pions and protons at rapidities y=0 and y$\approx$3.  
It is observed that the 
transverse energy decreases in going to forward rapidity, suggesting a possible reduction in the radial flow. This
change in the radial flow can strongly influence the differential elliptic flow behavior, but is expected to have
less of an effect on the integral flow.

In conclusion, BRAHMS has measured identified particle $v_2(p_T)$ at $\eta\approx$0, 1, and 3 for the Au+Au
system at $\sqrt{s_{\rm{NN}}}=200$ GeV.   The differential elliptic flow decreases at forward pseudorapidity, with the
decrease for central events consistent with the expectations of 3D Hydro+Cascade calculations.  For mid-central
collisions at forward pseudorapidity the elliptic flow is found to be significantly less than expected by hydrodynamic
calculations.   This reduction in the elliptic flow is also evident when the current results are compared to 
previous mid-rapidity results using constituent quark scaling.   A decrease is observed in the mean transverse
energy of particles going to forward pseudorapidity, suggesting a reduction in the radial flow component.   This change
in radial flow has a significant influence on the differential $v_2(p_T)$ values.  
In general, the forward pseudorapidity elliptic and
radial flow results place significant constraints on rapidity dependent model calculations of the dynamics of 
RHIC collisions.



\end{document}